\documentclass[12pt,preprint]{aastex}

\shorttitle{Full-Fledged Dwarf Irregular Galaxy Leo\,A}
\shortauthors{Vansevi\v{c}ius et al.}

\begin{document}

\title{Full-Fledged Dwarf Irregular Galaxy Leo\,A\altaffilmark{1}}


\author{Vladas Vansevi\v{c}ius\altaffilmark{2,3}, Nobuo
Arimoto\altaffilmark{2}, Takashi Hasegawa\altaffilmark{4}, Chisato
Ikuta\altaffilmark{2}, \\ Pascale Jablonka\altaffilmark{5},
Donatas Narbutis\altaffilmark{3}, Kouji Ohta\altaffilmark{6}, Rima
Stonkut\.{e}\altaffilmark{3}, \\ Naoyuki Tamura\altaffilmark{7},
Valdas Vansevi\v{c}ius\altaffilmark{3}, and Yoshihiko
Yamada\altaffilmark{2,8}}

\altaffiltext{1}{Based on data collected at Subaru Telescope,
which is operated by the National Astronomical Observatory of
Japan} \altaffiltext{2}{National Astronomical Observatory, Mitaka,
Tokyo 181-8588, Japan} \altaffiltext{3}{Institute of Physics,
Savanori\c{u} 231, Vilnius LT-03154, Lithuania}
\altaffiltext{4}{Gunma Astronomical Observatory, Gunma 377-0702,
Japan} \altaffiltext{5}{UMR 8111 CNRS, GEPI, Observatoire de
Paris, 92195 Meudon, France} \altaffiltext{6}{Department of
Astronomy, Kyoto University, Kyoto 606-8502, Japan}
\altaffiltext{7}{Department of Physics, University of Durham,
South Road, Durham, DH1 3LE, UK} \altaffiltext{8}{Institute of
Astronomy, University of Tokyo, Mitaka, Tokyo 181-0015, Japan}

\begin{abstract}
We have studied Leo\,A -- the isolated and extremely gas rich
dwarf irregular galaxy of very low stellar mass and metallicity.
Ages of the stellar populations in Leo\,A are ranging from
$\sim$10\,Myr to $\sim$10\,Gyr. Here we report the discovery of an
old stellar halo and a sharp stellar edge. Also we find the
distribution of stars extending beyond the gaseous envelope of the
galaxy. Therefore, Leo\,A by its structure as well as stellar and
gaseous content is found to resemble massive disk galaxies. This
implies that galaxies of very low stellar mass are also able to
develop complex structures, challenging contemporary understanding
of galaxy evolution.
\end{abstract}


\section{Introduction}

Understanding galaxy formation and evolution on the Hubble
timescale is one of astronomy's greatest challenges. The
importance of dwarf irregular galaxies (DIGs) and especially
stellar populations in their outskirts for study of the galaxy
build-up and star formation processes is well recognized and has
been widely discussed recently, see e.g., \citet{hod03}. DIGs are
excellent targets to study galaxy evolution, as their long
relaxation time makes them keep traces of interaction, merging
event or starburst at least for a few billion years. Many studies
\citep{min96,apa00,hel01,dro03} were devoted to search for the
outer stellar edges and extended old stellar populations in DIGs.
The radial reddening of the stellar populations distributed in the
disc-like structures was reported, however, the prominent edges
were never revealed \citep{hod03}.

For our study we selected the nearby DIG Leo\,A (Fig. 1). Leo\,A
is one of the most isolated galaxies in the Local Group. It is
extremely gas rich \citep{you96}, and possesses very low stellar
mass \citep{mat98} and metallicity \citep{ski89,vze99}. Basing on
the Hubble Space Telescope (HST) photometric observations in the
central part of the galaxy young age was deduced \citep{tol98}.
This conclusion challenged further investigations. Deep HST
photometry was performed in the adjacent field, and the signatures
of an old stellar population were identified \citep{sch02}. The
final proof of the old stellar population in Leo\,A was brought by
the discovery of the RR Lyr variables \citep{dol02}. Additionally,
it was noticed that the red giant branch (RGB) stars are
distributed more widely than the blue stars \citep{dol02}. Gaseous
content of the galaxy was investigated in detail \citep{you96},
and a symmetrical \ion{H}{1} envelope twice more extended than the
Leo\,A stellar distribution was discovered. Very small
lopsidedness measured in the red continuum and in the H$\alpha$
emission \citep{hel00} suggests that Leo A has not experienced any
strong event of merger or interaction for at least a few billion
years, and it is a good target for study of quiescent galaxy
evolution.

\section{Observations and Data Reductions}

Taking into account the angular size of Leo\,A (the Holmberg's
dimension $\sim$7$\arcmin\times$5$\arcmin$) \citep{mat98}, Subaru
Telescope equipped with Prime Focus Camera (Suprime-Cam)
\citep{miy02} is an ideal instrument to study stellar content at
the galaxy's very outskirts (Fig. 1). Single shot Suprime-Cam
mosaic (5$\times$2 CCD chips) covers a field of
34$\arcmin\times$27$\arcmin$ (pixel size
$0\farcs2\times0\farcs2$), and the magnitude of $V\sim25^m$ is
reached in 60\,s. We acquired images in three broad-band filters:
$B$ (5$\times$600\,s), $V$ (5$\times$360\,s) \& $I$
(30$\times$240\,s) during two photometric nights (seeing
$<$0$\farcs8$) on 2001 November 20-21. Standard reduction
procedures were performed with a software package \citep{yag02}
dedicated to the Suprime-Cam data. We employed six central CCD
frames of the mosaic, and performed crowded-field stellar
photometry by applying DAOPHOT \citep{ste87} implemented in the
IRAF software package \citep{tod93} on 5 individual exposures in
each photometric band. In order to use HST data for the central
part of the galaxy consistently, we transformed instrumental
magnitudes to the HST photometric system, $F439W$, $F555W$,
$F814W$ (for abbreviation further we use $B$, $V$, $I$) by
referring to the HST photometric data archive \citep{hol04}. The
transformation accuracy of $\sim$0$\fm01$ in $V$- \& $I$-bands,
and of $\sim$0$\fm02$ in $B$-band were achieved.

\section{Results}

In order to trace the entire extent of the old stellar populations
in Leo\,A we employed the RGB stars. The following RGB star
selection criteria have been applied: 1) location of stars in the
color-magnitude diagram (CMD), $I$ vs. ($V-I$), within the zone
marked in Fig. 2; 2) high photometric accuracy, $\sigma_I<0\fm06$
\& $\sigma_V<0\fm08$; 3) good fit with the stellar point spread
function (DAOPHOT parameters: $\chi^2$$<1.5$;
$\vert$sharpness$\vert<0.4$); 4) photometric criterion devoted to
wipe out the objects with non-stellar spectra, $(B-V)-(V-I)$, in
the range of ($-$0.40) -- (+0.10). The main parameters deduced for
the RGB star distribution are: center at RA(2000) =
$09^h59^m24\fs0$, Dec(2000) = 30\degr 44\arcmin 47\arcsec;
ellipticity (ratio of semi-minor to semi-major axis) b/a =
0.60$\pm$0.03 coinciding with b/a of the \ion{H}{1} envelope
\citep{you96}; position angle of the major axis PA =
114\degr$\pm$5\degr. For detailed examination we selected the
field located inside the ellipse (b/a = 0.60) of a = 12\arcmin\,
centered at the derived position (stellar content is shown in Fig.
2a), which is large enough to comfortably accommodate Leo\,A
inside. We detected 1394 RGB stars distributed {\it symmetrically
and smoothly} within this field. The average accuracies for the
RGB star sample are: 1) coordinate matching among the $B$-, $V$-
\& $I$-bands, 0\farcs06; 2) photometry, $\sigma_B$ = 0.019 (4.6),
$\sigma_V$ = 0.015 (4.7), $\sigma_I$ = 0.010 (4.7) (the average
number of detections in each band is given in brackets).

The radial profile (RP) of the RGB star surface number density
[arcmin$^{-2}$] (Fig. 3) was constructed by integrating within
elliptical (b/a = 0.60) rings of a width, $\Delta$a = 0\farcm5 (at
the Leo\,A distance, 800\,kpc, 1\arcmin\, corresponds to 230 pc).
We performed RP robustness tests by varying b/a = (0.55, 0.60,
0.65), PA = (104\degr, 114\degr, 124\degr), and $\Delta$a =
0\farcm2 -- 1\farcm0, as well as a magnitude at the faint limit,
$I$ = (22.5, 23.0, 23.5), and found no significant change in the
RP's form. Five distinct RP zones are noticeable (Fig. 3): 1)
crowded central part, a = 0\farcm0 -- 2\farcm0 (completeness at $I
= 23^m$ varies with radius from 80 to 90\%); 2) old exponential
disk extending far beyond the previously estimated size of the
galaxy \citep{mat98}, a = 2\farcm0 -- 5\farcm5 (for representative
stellar content see Fig. 2b), scale-length (S-L)
1\farcm03$\pm$0\farcm03; 3) discovered new stellar component in
DIGs, which we call ``halo'', a = 5\farcm5 -- 7\farcm5 (for
stellar content see Fig. 2c), S-L 1\farcm84$\pm$0\farcm09; 4)
sharp cut-off of the RGB star distribution coinciding with the
observed edge \citep{you96} and predicted cut-off of the
\ion{H}{1} envelope \citep{ste02}, a = 7\farcm5 -- 8\farcm0; 5)
sky background zone where we derived a number density of
contaminants to the RGB stars, a = 8\farcm0 -- 12\farcm0 (for
representative stellar content see Fig. 2d). In order to show the
entire extent of the discovered structures in Leo\,A we
over-plotted some characteristic size ellipses (b/a = 0.60) on the
Suprime-Cam $V$-band image (Fig. 1). It is worth noting that the
suspected Leo\,A RR Lyr variable C1-V01 \citep{dol02} is located
just outside the ellipse marking the galaxy size, confirming old
age and large extent of the discovered stellar halo.

\section{Discussion and Conclusion}

In the Cold Dark Matter (CDM) cosmology scenarios galaxies are
assumed to build up and develop their internal structure via
hierarchical merging of the primordial amplified density
fluctuations into larger systems. Therefore, our discovery of the
stellar populations possessing distinct spatial distributions in
the undisturbed very low mass DIG, Leo\,A, which is unlikely built
via merging, suggests an alternative way of galaxy structure
formation. In order to quantify the discovered stellar populations
we estimate some of their basic parameters.

From the determined RP of the RGB stars it is straightforward to
evaluate the total stellar mass of the old stellar population in
Leo\,A. Assuming the Salpeter's initial mass function and a
stellar mass range of 0.5 -- 100 M$_{\sun}$, we derived the mass
of $\sim$4$\cdot$10$^6$ M$_{\sun}$ (an isochrone by \citet{gir02}
of 10\,Gyr and Z = 0.0004 has been employed). This is in agreement
with the recent estimate \citep{lee03} and confirms very low mass
of the stellar populations in Leo\,A.

We performed a halo-disk RP decomposition considering two extreme
cases: 1) when the exponential disk, S-L 1\farcm03, is subtracted
as the primary population, the remaining halo mass is $\sim$3\% of
the disk mass; 2) when the exponential halo, S-L 1\farcm84, is
assumed to be a primary population, the halo mass is $\sim$30\% of
the disk mass. The estimated lower and upper mass fractions of the
halo are comparable to the Milky Way's halo and thick disk cases
\citep{rob03}, respectively. The young ($<$1 Gyr) disk population
(stars located in CMD, $I<24^m$ \& $(V-I)<0.25$, Fig. 2a) in
Leo\,A is traceable up to the radius of a $\sim$5\arcmin, and is
exponentially distributed in the disk of S-L 0\farcm56$\pm$0.06.
The spatial distributions of the RGB stars and gas \citep{you96}
possess sharp coincident cut-offs at large radius, implying that
the disk properties of DIGs and of the massive galaxies
\citep{vdk01} are similar.

We conclude, that the young and old Leo\,A disks together with the
discovered old halo and sharp stellar edge closely resemble basic
structures found in the large full-fledged disk galaxies. This
suggests complex formation histories even in the very low stellar
mass galaxies like Leo\,A, and challenges contemporary
understanding of galaxy evolution.

\acknowledgments

We thank Alvio Renzini for a fruitful discussion and Jon Holtzman
for advices on the use of his HST photometric data archive. V.V.
acknowledges the National Astronomical Observatory of Japan for a
professorship. This work was financially supported in part by a
Grant-in-Aid for Scientific Research by the Japanese Ministry of
Education, Culture, Sports, Science and Technology (No. 13640230),
and by a Grant T-67/04 of the Lithuanian State Science and Studies
Foundation.

\clearpage


\begin{figure}
\epsscale{1.0} \plotone{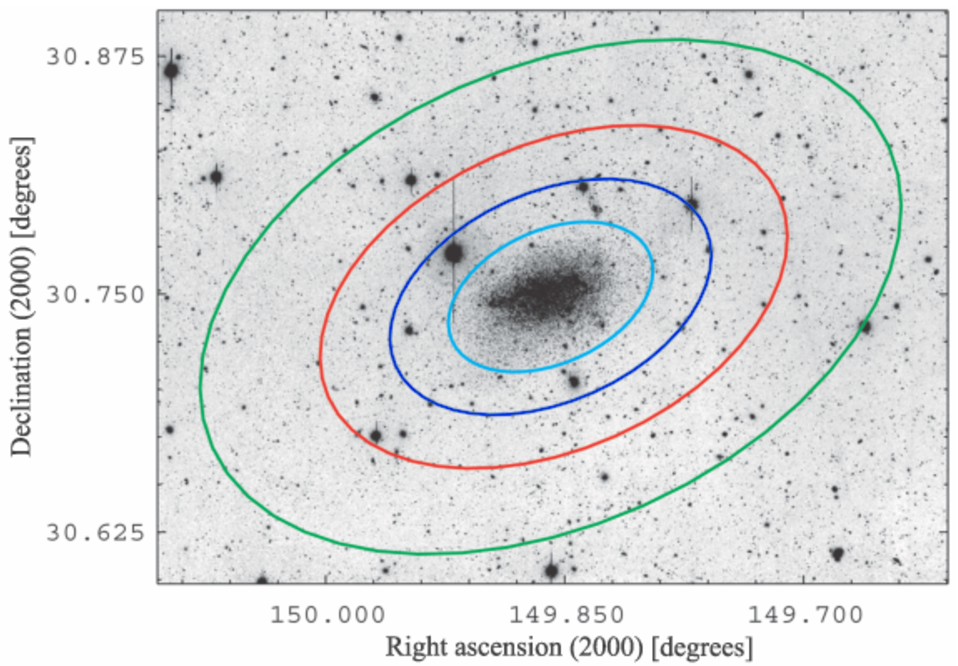} \caption{The Suprime-Cam $V$-band
image of the galaxy Leo\,A. The ellipses (b/a = 0.6, a -
semi-major axis) indicate: the Holmberg's radius ($B = 26\fm5$
arcsec$^{-2}$), a = 3\farcm5 (cyan); the radial distance where the
discovered halo becomes prominent, a = 5\farcm5 (blue); the size
of Leo\,A established in this work, a = 8\farcm0 (red); the zone
used for background source surface number density determination, a
= 12\farcm0 (green).\label{fig1}}
\end{figure}

\clearpage

\begin{figure}
\epsscale{1.0} \plotone{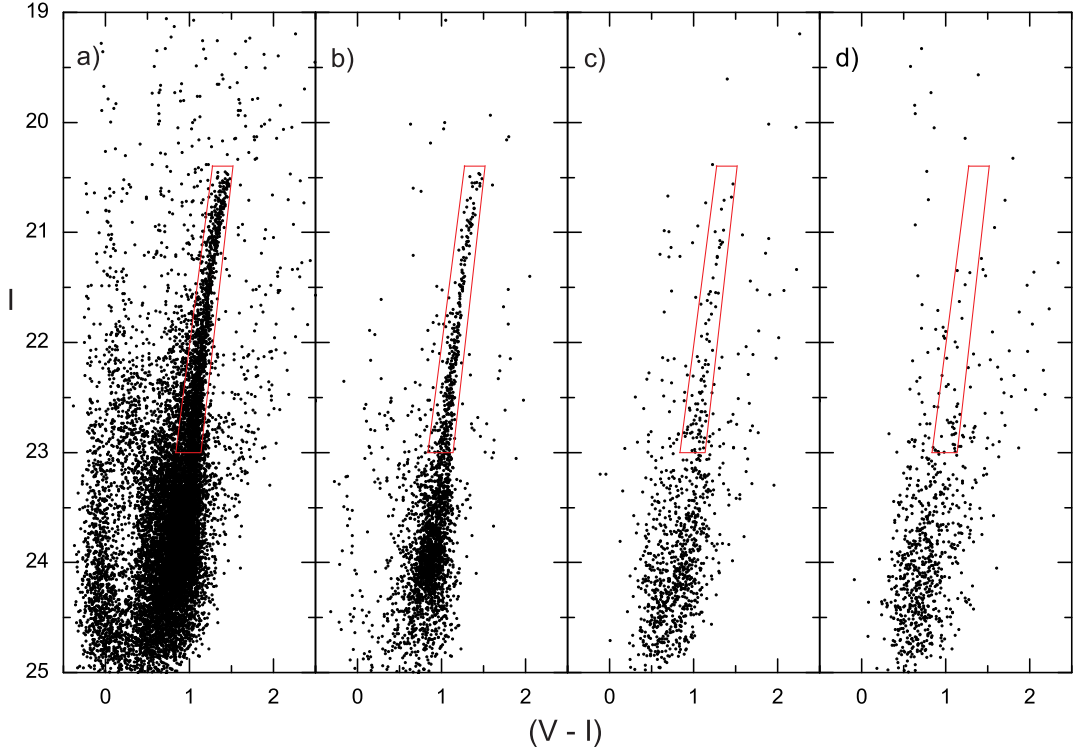} \caption{The color-magnitude
diagrams of the stellar-like objects in Leo\,A. The objects are
shown: (a) in the elliptical (b/a = 0.6) area, a $<$ 12\arcmin,
containing Leo\,A and its surroundings, the number of objects
plotted, N = 12,604; (b) in the representative old disk area,
3\farcm0 $<$ a $<$ 5\farcm0, N = 2,462; (c) in the discovered halo
area, 5\farcm5 $<$ a $<$ 7\farcm5, N = 974; (d) in the background
area, 9\farcm0 $<$ a $<$ 10\farcm5, N = 780. The RGB stars
employed for the structure analysis of Leo\,A were selected from
the zone marked by lines: magnitude ranges from the RGB tip $I =
20\fm4$ down to $I = 23\fm0$; the inclined lines are given by the
equations $I=31-7\cdot(V-I)$ and $I=28-6\cdot(V-I)$.\label{fig2}}
\end{figure}

\clearpage

\begin{figure}
\epsscale{1.0} \plotone{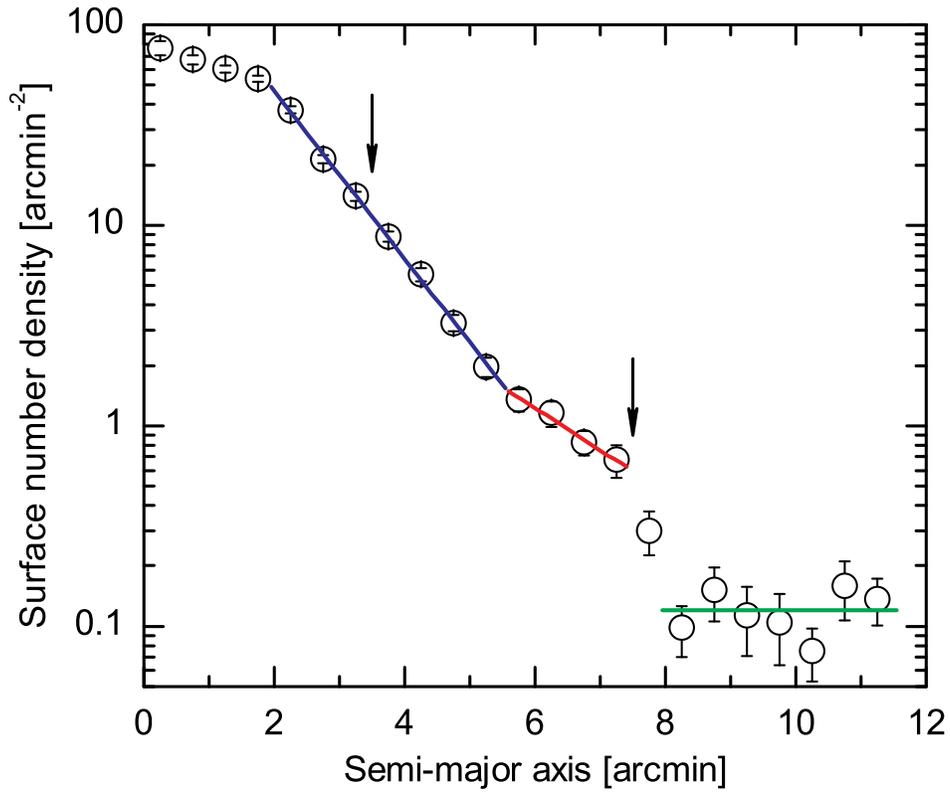} \caption{The radial profile of the
RGB star surface number density in Leo\,A. The lines fitted to the
old disk, 2\farcm0 $<$ a $<$ 5\farcm5, and the halo, 5\farcm5 $<$
a $<$ 7\farcm5, radial profiles, and the background, 8\farcm0 $<$
a $<$ 12\farcm0 are shown. The Holmberg's radius (a = 3\farcm5)
and the observed size of the \ion{H}{1} envelope \citep{you96} (a
= 7\farcm5) are indicated with arrows.\label{fig3}}
\end{figure}

\end{document}